\documentclass[12pt]{article}
\usepackage[margin=1in]{geometry}
\usepackage{amsmath}
\usepackage{amsthm}
\usepackage{amssymb}
\usepackage{bm}
\usepackage{bbm,mathrsfs}
\usepackage{graphicx}
\usepackage{color}
\usepackage[]{hyperref}

\numberwithin{equation}{section}
\def\mysavedown#1{\edef\mysubs{\mysubs#1}}
\def\mysaveup#1{\edef\mysups{\mysups#1}}
\def\mydown#1{{\mytensor}_{\vphantom{\mysubs}#1}}
\def\myup#1{{\mytensor}^{\vphantom{\mysups}#1}}
\def\tensor#1#2{
  #1
  \def\mytensor{\vphantom{#1}}
  \def\mysubs{\relax}
  \def\mysups{\relax}
  \let\down=\mysavedown
  \let\up=\mysaveup
  #2
  \let\down=\mydown
  \let\up=\myup
  #2
  }

\DeclareMathOperator{\Tr}{Tr}

\newcommand{\Atil}{\widetilde{A}}
\newcommand{\Btil}{\widetilde{B}}

\newcommand{\Htil}{\widetilde{H}}
\newcommand{\Hden}{\mathcal{H}}
\newcommand{\Mtil}{\widetilde{M}}

\newcommand{\Phitil}{\widetilde{\Phi}}

\newcommand{\Ttil}{\widetilde{T}}

\newcommand{\Vtil}{\widetilde{V}}
\newcommand{\Wtil}{\widetilde{W}}

\newcommand{\gtil}{\tilde{g}}

\newcommand{\half}{\frac{1}{2}\;}

\newcommand{\ltil}{\tilde{l}}

\newcommand{\mus}{\mu_\sigma}
\newcommand{\mut}{\mu_\tau}
\newcommand{\mtil}{\tilde{m}}

\newcommand{\ntil}{\tilde{n}}

\newcommand{\omegatil}{\tilde{\omega}}

\newcommand{\ptil}{\tilde{p}}
\newcommand{\qtil}{\tilde{q}}

\newcommand{\util}{\tilde{u}}

\newcommand{\xstil}[1]{\xtil^{#1}{}_{\sigma}}
\newcommand{\xs}[1]{x^{#1}{}_{\sigma}}

\newcommand{\xtil}{\tilde{x}}

\begin{document}
\begin{titlepage}
\strut\mbox{December 2007}
\par\vskip 2cm
\begin{center}
    {\Large \bf Target Space Duality: The Dilaton Field}\\[.3in]
    %author
    
    {\bf Orlando Alvarez}\footnote{email: \tt oalvarez@miami.edu}\\
    {\bf B\l a\.{z}ej Ruszczycki}\footnote{email: \tt ruszczycki@physics.miami.edu}\\[0.1in]
    %address
    Department of Physics\\
    University of Miami\\
    P.O. Box 248046\\
    Coral Gables, FL 33124 USA\\[0.3in]
\end{center}
\par\strut\vspace{.5in}

%abstract
\begin{abstract}
Classical target space duality transformations are studied for the
non-linear sigma model with a dilaton field.  Working within the
framework of the  Hamiltonian formalism we require the duality
transformation to be a property only of the target spaces.  We obtain
a set of restrictions on the geometrical data.  The ``on-shell duality''
integrability conditions are inspected.
\end{abstract}

\vspace{.25in}


\end{titlepage}

\section{Introduction}
\label{sec:introduction}
In a series of previous papers \cite{Alvarez:2000bh,Alvarez:2000bi} a
framework was developed for studying classical target space duality
between nonlinear sigma models in two dimensional Minkowski space.
References to the earlier literature may be found in
\cite{Alvarez:2000bh,Alvarez:2000bi}.  Here we introduce a dilaton
field $\Phi$ coupled to the world sheet curvature scalar
$R^{(2)}(\vec{\sigma})$ via the action $S=S_0+S_B+S_\Phi$ where
\begin{equation}
S_0+S_B=-\half\int\ d^2\sigma \{ \sqrt{-h}h^{\alpha\beta}g_{ij}(x)\partial_\alpha x^i \partial_\beta x^j
-\epsilon^{\alpha\beta}\partial_\alpha x^i  \partial_\beta x^j B_{ij}(x) \},
\end{equation}
\begin{equation}
S_\Phi=\int\ d^2\sigma \sqrt{-h} \Phi(x)R^{(2)}(\vec{\sigma}).
\end{equation}
The target manifolds with their respective geometrical data are
denoted by $M(g,B,\Phi)$ and $\Mtil(\gtil,\Btil,\Phitil)$.  The Greek
indices refer to the world sheet $\Sigma$ with metric
$h_{\alpha\beta}$ and coordinates $\vec{\sigma}=(\tau,\sigma)$.  In two dimensions it is always possible to find a
coordinate transformation that locally puts the metric in conformal
form
\begin{equation}
h_{\alpha\beta}=e^{2\mu(\vec{\sigma})}\eta_{\alpha\beta},\label{wsmetric}
\end{equation}
where $\eta_{\alpha\beta}$ is the flat world sheet metric with the
signature $(-,+)$ and $\mu(\vec{\sigma})$ is the conformal factor.
Introducing light-cone coordinates on the world sheet by
$\sigma^{\pm}=\tau \pm \sigma$ we see that the curvature scalar is
\begin{equation}
R^{(2)}=8e^{-2\mu(\vec{\sigma})}\;\partial^{2}_{+-}\mu(\vec{\sigma}).
\end{equation}

A possible multiplicative constant for $S_\Phi$ can be absorbed into
the definition of dilaton field $\Phi$.  Defining the closed 3-form
$H$ by $H=dB$ and the derivatives $x^{i}_{\pm}$ by pulling back an
orthonormal coframe from the target space\footnote{See
\cite{Alvarez:2000bi,Alvarez:2000pk} for notation.},
$\theta^{i}=x^{i}{}_{\alpha}\,d\sigma^{\alpha}$ gives the classical
equations of motion
\begin{equation}
x^i{}_{+-}+\half H^i{}_{jk}(x)\;x^j_+x^k_-+2\;
\Phi'_{i}(x)\;\partial^{2}_{+-}\mu(\vec{\sigma})=0\,\label{eom}
\end{equation}
where $d\Phi = \Phi'_{i}\,\theta^{i}$ and $x^{i}{}_{+-}$ is the 
second covariant derivative.
There is an analogous expression for $\Mtil$ model.  Any
transformation of the metric resulting from
\begin{equation}
\mu(\vec{\sigma}) \rightarrow \mu(\vec{\sigma})
+\xi(\sigma^+)+\eta(\sigma^-)\label{gauging}
\end{equation}
with arbitrary functions $\xi$ and $\eta$ leaves the form of
equations of motion invariant.

We note that the first two terms in the action $S_0+S_B$
are manifestly independent of the choice of the conformal factor but
the term $S_\Phi$ is not\footnote {It is still invariant under
simultaneous global scaling of the conformal factor $\mu(\vec{\sigma})
\rightarrow a\, \mu(\vec{\sigma})$ and the dilaton field $\Phi(x)
\rightarrow \Phi(x)/a$.}.  

The $\Phi=0$ models are trivially classically conformally invariant.
Therefore, at the classical level it was sufficient to study those
models on a flat world sheet because all dependence on the conformal
factor was absent.  Consequently, the study of classical target space
duality in these conformally invariant models reduced to studying
models with a flat world sheet.  The classical conformal invariance is
manifestly broken by the presence of a generic non-zero dilaton
field\footnote {At the quantum level it is consistent to choose the
dilaton term to be $O(\hbar)$.  The condition for conformal
invariance, the tracelessness of the energy-momentum tensor, is
obtained by combining contributions (both classical and quantum) from
the dilaton term with the quantum corrections from the other terms.
This gives a set of restrictions on geometrical data describing a
conformally invariant model, see e.g.
\cite{GSW:Strings2,Polchinski:Strings1}.}.  This means that the
classical behavior of strings propagating on a target space $M$
depends on local metrical properties of the world sheet $\Sigma$.  In
this article we ask the following question.
\begin{quote}
    Is it possible to have a classical duality transformations between
    strings propagating on target spaces $M$ and $\Mtil$ such that the
    duality transformation is only a property of the target spaces and
    it is independent of the metrical geometry of the world sheet
    $\Sigma$, \emph{i.e.}, independent of the conformal factor $\mu$?
\end{quote}
N.B. The sigma models are not necessarily conformally invariant.

The world sheet stress-energy tensor $T^{\alpha\beta}$ is defined 
variationally by
\begin{equation}
\delta S=\half\int  d^2\sigma  \sqrt{-h} T^{\alpha\beta} \delta h_{\alpha\beta}.
\end{equation}
The contribution from $S_0$ is  
\begin{equation}
T_{(0)}^{\alpha\beta}=4e^{-4\mu(\vec{\sigma})}
\begin{pmatrix}
x_-\; x_- & 0\\
0 & x_+ \;  x_+\
\end{pmatrix}, 
\end{equation}
the term $S_B$ does not contribute to the stress-energy tensor
although it contribute to the equations of motion.  Using the
equations of motion in the calculation of $\nabla_\alpha
T^{\alpha\beta}$ you find terms such as $H_{ijk}x^i_-x^j_-x^k_+$ that
vanish due to the antisymmetry of $H_{ijk}$.

To calculate the contribution from $S_\Phi$ we have to integrate by
parts twice and to observe that the Einstein tensor vanishes because
the Hilbert-Einstein action is a topological invariant in $2$
dimensions.  After some algebra we arrive at the result
\begin{equation}
T_{(\Phi)}^{\alpha\beta}=2(h^{\gamma\alpha}h^{\delta\beta}\Phi_{\gamma;\delta}-h^{\alpha\beta}
h^{\gamma\delta} \Phi_{\gamma;\delta}),
\end{equation}
where $ \Phi_ \gamma :=x^i_{\gamma}\, \partial_i \Phi(x) $ and $\Phi_{
\gamma; \delta} $ is the covariant derivative on the world sheet.  The
only non-vanishing connection coefficients for the metric
\eqref{wsmetric} are
$\Gamma^+_{++}=2\partial_+\mu(\vec{\sigma})$ and
\mbox{$\Gamma^-_{--}=2\partial_-\mu(\vec{\sigma})$.} The
explicit expression is
\begin{equation}
T_{(\Phi)}^{\alpha\beta}=8e^{-4\mu}
\begin{pmatrix}
2\, \partial_-\mu \; \Phi'_j \, x^j_- - \Phi'_{i;j}\,x^i_- x^j_- - \Phi'_i \,x^i_{--}
& \Phi'_{i;j}\,x^i_+x^j_-+\Phi'_i \,x^i_{+-}\\
\Phi'_{i;j}\,x^i_+x^j_-+\Phi'_i \,x^i_{+-} & 2\, \partial_+\mu \, \Phi'_j \, x^j_+ - \Phi'_{i;j}\,x^i_+ x^j_+ - \Phi'_i \,x^i_{++}
\end{pmatrix}. 
\end{equation}
Here, the derivatives $\Phi'_i$ are defined as components of the 1-form $d\Phi$ with respect
to the orthonormal coframe of $M$ and $\Phi'_{i;j}$ is a covariant derivative on the target space. 
For the model $\Mtil(\gtil,\Btil,\Phitil)$ these derivatives will be denoted as $\Phitil''_i$ and $\Phitil''_{i;j}$ respectively.
Note that the stress-energy tensor does not vanish in the limiting case of a flat world
sheet $\mu(\vec{\sigma})\rightarrow0$, although in this limit both the
action and the classical equations of motion are the same as for the
flat world sheet case.  In the limit $\mu(\vec{\sigma})\rightarrow0$,
$T_{(\Phi)}^{\alpha\beta}$ has the property that its divergence
vanishes identically, i.e. $\nabla_\beta T_{(\Phi)}^{\alpha\beta}=0$
for any $x^i(\sigma^\mu)$ and not just for the solution of equation of
motion.

In the case of \cite{Alvarez:2000bh,Alvarez:2000bi} where the action
was $S_0+S_B$, the transformation equation for the ``on-shell'' duality
could have been written down by inspecting
the stress-energy tensor.  Here there are two difficulties. The
stress-energy tensor contains the terms $\Phi'_i(x)$ and
$\Phi'_{i;j}(x)$ which have dependence on $x$'s and the duality
transformation involves derivatives of $x$'s,
therefore  integrability issues arise.  The other difficulty
is just the mentioned possibility that two equivalent expressions may
differ by a contribution whose divergence vanishes identically and this
contribution has to be clearly identified.
 
\section{A Toy Model}
\label{sec:toymodel} 

As a guideline consider a classical mechanics time-dependent
Hamiltonian system.  By prescribing a generating function
$F(q,\qtil,t)$ one obtains both the canonical transformation and the 
relation between the hamiltonians by
\begin{equation}
\ptil=\frac{\partial F(q,\qtil,t)}{\partial \qtil}, \qquad
-p=\frac{\partial F(q,\qtil,t)}{\partial q}, \qquad
\Htil-H=-\frac{\partial F(q,\qtil,t)}{\partial t} \,.
\label {ctrans}
\end{equation}
Here we consider the inverse problem.  Both hamiltonians are given and
we want to determine the conditions that have to be satisfied in
order to establish a canonical transformation.  We assume the 
Hamiltonians are of the form
\begin{equation}
H=\half (p-A(q,t))^2+V(q,t), \qquad \Htil=\half (\ptil-\Atil(\qtil,t))^2+\Vtil(\qtil,t).\label{cmhams}
\end{equation}
Consider a generating function of the form
\begin{equation}
F(q,\qtil,t)=q\qtil+f(t)(\Wtil(\qtil)-W(q))
\end{equation}
with $f(t),\Wtil(\qtil),W(q)$ to be determined.
Using \eqref{ctrans} we see that
\begin{equation}
q=\ptil-f \Wtil', \qquad \qtil=-(p-fW')\label{ctrans1}
\end{equation}
and 
\begin{equation}
\half(\ptil-\Atil)^2+\Vtil- \left\{\half(p-A)^2+V
\right\}=-\dot{f}(\Wtil-W).
\label{ctransH}
\end{equation}
We rewrite \eqref{ctransH} using \eqref{ctrans1} and group together terms 
according to their  $q$ and $\qtil$  dependence:
\begin{align}
0=& \half q^2-V-\half(A-fW')-\dot{f}W \notag \\
 &\quad -\left\{\half \qtil^2-\Vtil-\half(\Atil-f\Wtil')-\dot{f}\Wtil
 \right\} \notag\\
 &\quad -q(\Atil-f\Wtil')-\qtil(A-fW').\label{cm_con}
\end{align}
To eliminate the mixed $q$ and $\qtil$ dependence we require that the
summands in the last line of the equation above satisfy
\begin{equation}\label{cm_conf}
A-fW'=h(t)q, \qquad \Atil-f\Wtil' =-h(t)\qtil.
\end{equation}
The remaining part of \eqref{cm_con} gives immediately the conditions
\begin{align}
V(q,t)=\half(1-h(t))^2q^2-\dot{f}W,\notag\\
\Vtil(\qtil,t)=\half(1-h(t))^2\qtil^2-\dot{f}\Wtil.
\end{align}

To make this example more similar to the sigma model case consider 
the special case where
\begin{equation}
A(q,t)=f(t)B(q),\qquad A(\qtil,t)=f(t)\Btil(\qtil).
\end{equation}
From \eqref{cm_conf} we see that
\begin{equation}
B-W'=q , \qquad \Btil-\Wtil'=\qtil,\qquad h(t)=f(t).
\end{equation}
Integrating we obtain the generating function for this transformation
\begin{equation}
F(q,\qtil,t)=q\qtil+h(t)\left(-\half(\qtil^2-q^2)
+\int_0^{\qtil} d \qtil' \,\Btil(\qtil')-\int_0^{q} d q' \,B(q')\right).
\end{equation}

\section{Target Space Duality}
\label{sec:target-space-duality}

In the  field theory case we have to consider  hamiltonian densities 
of the form
\begin{equation}
\Hden=\half g^{ik} (\pi_i-B_{ij}x^j_\sigma)(\pi_k-B_{kl}x^l_\sigma)+g_{ik}\half x^i_\sigma x^k_\sigma+2\eta^{\alpha\beta}(\partial^2_{\alpha\beta}\mu)\Phi(x)
\end{equation}
and the analogous expression for the $\Mtil$ model.  The explicit time
dependence enters via the conformal factor.  By analogy to
\eqref{ctrans} the imposed requirement is that the Hamiltonians of
both models differ only by a time derivative of the generating
functional.
\begin{equation}
 \Htil-H=\int d\sigma \;  (\widetilde{\Hden}- \Hden)=-\frac{\partial F}{\partial{ \tau}}\label{hdenscon}
\end{equation}
The general form of the generating functional is taken to be of the 
form
\begin{equation}
F[x,\xtil]=\int \alpha + \int( \partial_\sigma \mu \; Y + 
\partial_\tau \mu \;Z)d \sigma,
\label{eq:h-1def}
\end{equation}
where  $Y(x,\xtil)$ and 
$Z(x,\xtil)$ are functions on $M\times \Mtil$ and $\alpha(x,\xtil)$ is a $1$-form on $M\times \Mtil$ written as
\begin{equation}
\alpha=\alpha_i(x,\xtil)dx^i+\tilde{\alpha}_i(x,\xtil)d\xtil^i.
\end{equation}
The associated canonical transformation is 
\begin{align}
\pi_i&=m_{ji}\; \frac{d \xtil^j}{d\sigma}+l_{ij}\;\frac{dx^j}{d\sigma}- \partial_\sigma \mu \;\frac{\partial Y}{\partial x^i}- \partial_\tau \mu \;\frac{\partial Z}{\partial x^i},\\
\widetilde{\pi}_i&=m_{ij} \;\frac{d x^j}{d\sigma}+\ltil_{ij}\;\frac{d\xtil^j}{d\sigma}+ \partial_\sigma \mu \;\frac{\partial Y}{\partial \xtil^i}+ \partial_\tau \mu \;\frac{\partial Z}{\partial \xtil^i},\label{cantransq}
\end{align}
where $m_{ij}$, $l_{ij}$ and $\ltil_{ij}$ are given by 
\begin{equation}
d\alpha=-\half l_{ij}(x,\xtil) \,dx^i\wedge dx^j+\half \ltil_{ij}(x,\xtil) \,d\xtil^i\wedge d\xtil^j+
m_{ij}(x,\xtil) \,d\xtil^i\wedge dx^j.
\end{equation}

Here it would be desirable to maintain a symmetric formulation
between tilded and untilded quantities.  Therefore we introduce the
following definitions:
\begin{equation}
n\equiv l-B, \qquad\qquad \ntil\equiv \ltil-\widetilde{B},
\end{equation}
\begin{equation}
\widetilde{m}_{ij}\equiv m_{ji},\label{mdef}
\end{equation}
\begin{align}
dY&=Y'_i\;\theta^i-Y''_i\;\widetilde{\theta}^i,\\
dZ&=Z'_i\;\theta^i-Z''_i\;\widetilde{\theta}^i, 
\end{align}
where ($\theta$,$\widetilde{\theta}$) is an orthonormal coframe of $M\times\Mtil$.
We use $\| \; \|$ and $\langle \;, \; \rangle$ to denote the norms 
and the 
inner products on the target spaces, we also suppress target space
indices $i,j \dots$ hereafter.  Using the form of the canonical
transformations we see that the integrand of \eqref{hdenscon} is
\begin{align}
 \widetilde{\Hden}- \Hden=\half &\| m x_\sigma+ \ntil \xtil_\sigma-\mus \;Y'' -\mut \;Z'' \|^2+ \half \| \xtil_\sigma\|^2+\notag \\
-&\| \mtil \xtil_\sigma+ n x_\sigma-\mus\;Y'-\mut \;Z' \|^2+ \half \| x_\sigma \|^2+\notag \\
-&2(-\mu_{\tau\tau}+\mu_{\sigma\sigma})(\Phitil-\Phi).
\end{align}
The next step is to group together terms with different $x$ and 
$\xtil$ behavior:
\begin{align}
 \widetilde{\Hden}- \Hden=&\half \langle x_\sigma,( m^t m-n^tn-I)\,x_\sigma\rangle-\half \langle\xtil_\sigma,( \mtil^t \mtil-\ntil^t\ntil-I)\,\xtil_\sigma\rangle\notag\\
 &+\langle \xtil_\sigma,( \ntil^t m- \mtil^t n)\,x_\sigma\rangle\notag\\
 &+\mut[-\langle x_\sigma,m^t \, Z'' -n^t \, Z'\rangle+\langle \xtil_\sigma,\mtil^t \, Z' -\ntil^t \, Z''\rangle]\notag\\
&+\mus[-\langle x_\sigma,m^t \, Y'' -n^t \, Y'\rangle+\langle \xtil_\sigma,\mtil^t \, Y' -\ntil^t \, Y''\rangle]\notag\\
 &+\half \| \mus \;Y'' +\mut \;Z'' \|^2-\half \| \mus \;Y' +\mut \;Z' \|^2\notag\\
 &-2(-\mu_{\tau\tau}+\mu_{\sigma\sigma})(\Phitil-\Phi ).\label{hdensgr}
\end{align}
On the level of Hamiltonian densities the condition \eqref{hdenscon} is expressed as 
\begin{align}
 \widetilde{\Hden}- \Hden&=-( \mu_ {\sigma
 \tau}\,Y+\mu_{\tau\tau}\,Z)+\frac{\partial}{\partial\sigma}h\notag.
\end{align} 
It will be convenient to have this condition rewritten as
\begin{align}
 \widetilde{\Hden}- \Hden&=\mut(Y'_i\,\xs i -Y''_i\,\xstil i) +\mus(Z'_i\,\xs i -Z''_i\, \xstil i)\notag\\
 &+(-\mu _{\tau \tau}+\mu _{\sigma \sigma})\,Z+ \frac{d}{d \sigma}\left(h-\mut\,Y-\mus\,Z \right).\label{hdensgr1}
\end{align}
Looking at $x_\sigma x_\sigma$, $\xtil_\sigma \xtil_\sigma$, $\xtil_\sigma x_\sigma$ terms in \eqref{hdensgr} 
we recover the relations
\begin{align}
\mtil^{t} \mtil & =I+\ntil^t \ntil=I- \ntil^{2}=mm^t\;, \label{eq:mmt}  \\    
    m^{t}m & = I+n^tn= I - n^{2}\;, \label{eq:mtm}  \\
    -mn & =  \ntil m, \label{eq:mn}
\end{align}
which are the same as the ones calculated in \cite{Alvarez:2000bh}.
Incorporating these in the remaining terms of \eqref{hdensgr} and using \eqref{hdensgr1} we
obtain
\begin{align}
\frac{d}{d\sigma}\left(h-\mut\,Y-\mus Z \right)
&=\mut[-\langle x_\sigma,m^t \, Z'' +n \, Z'+Y'\rangle+\langle 
\xtil_\sigma,\mtil^t \, Z' +\ntil \, Z''+Y''\rangle] \nonumber\\
&+\mus[-\langle x_\sigma,m^t \, Y'' +n \, Y'+Z'\rangle+\langle 
\xtil_\sigma,\mtil^t \, Y' +\ntil \, Y''+Z''\rangle] \nonumber \\
&+\half\| \mus \;Y'' +\mut \;Z'' \|^2-\half \| \mus \;Y' +\mut \;Z' 
\|^2 \nonumber \\
&+(\mu_{\tau\tau}-\mu_{\sigma\sigma})\,[2(\Phitil-\Phi )+Z].
\label{eq:h-2}
\end{align}
Now we require that our construction be independent of the conformal
factor $\mu$.  The terms linear in $\mut$ and $\mus$ should vanish
which gives us the equations:
\begin{equation}
\begin{pmatrix}
Y'\\
Y'' 
\end{pmatrix}=
-\begin{pmatrix}
n & m^t\\
\mtil^t & \ntil 
\end{pmatrix}
\begin{pmatrix}
Z'\\
Z''
\end{pmatrix},
\end{equation}

\begin{equation}
\begin{pmatrix}
Z'\\
Z'' 
\end{pmatrix}=
-\begin{pmatrix}
n & m^t\\
\mtil^t & \ntil 
\end{pmatrix}
\begin{pmatrix}
Y'\\
Y''
\end{pmatrix}.
\end{equation}
The above matrix equations are equivalent because
\begin{equation}
\begin{pmatrix}
n & m^t\\
\mtil^t & \ntil 
\end{pmatrix}^{2}=\mathbbm{1}.
\end{equation}
Next we  concentrate on
terms quadratic in first derivatives of the conformal factor.  They give us respectively the following equations:
\begin{align}
&\mu_\tau ^2:& &\|Z''\|^2-\|Z'\|^2=0,\\
&\mu_\sigma ^2:& &\|Y''\|^2-\|Y'\|^2=0,\\
&\mu_\tau \mu_\sigma:& &\langle Y'',Z''\rangle-\langle Y',Z' \rangle=0.
\end{align}
From the term linear in second-order derivatives we obtain that
\begin{equation}
Z=2(\Phitil-\Phi) .
\end{equation}
This gives
us immediately a condition
\begin{equation}
\lVert \Phi' \rVert^{2}= \lVert\Phitil''\rVert^{2}.\label{phicond}
\end{equation}
The L.H.S. of \eqref{phicond} is only a function of $x$, the R.H.S.
only of $\xtil$ which means that it is in fact a restriction saying
that the $1$-forms $d\Phi$ and $d\Phitil$ have the same norm in their
respective metrics.

Now we are ready to write down the duality equation  
% \begin{equation}
% \begin{pmatrix}
% \xtil_\tau+ 2 \mut \Phitil \\
% \xtil_\sigma+ 2 \mus \Phitil
% \end{pmatrix}=
% \begin{pmatrix}
% \ntil(m^t)^{-1} & m-\ntil (m^t)^{-1}n\\
% (m^t)^{-1} & -(m^t)^{-1}n 
% \end{pmatrix}
% \begin{pmatrix}
% x_\tau+ 2 \mut \Phi \\
% x_\sigma+ 2 \mus \Phi
% \end{pmatrix}. 
% \end{equation}
\begin{align}
\begin{pmatrix}
\xtil_\sigma+ 2 \mus \Phitil''\\
\xtil_\tau+ 2 \mut \Phitil'' 
\end{pmatrix} & =
\begin{pmatrix}
    -(m^t)^{-1}n &   (m^t)^{-1} \\
m-\ntil (m^t)^{-1}n & \ntil(m^t)^{-1}
\end{pmatrix}
\begin{pmatrix}
x_\sigma+ 2 \mus \Phi'\\
x_\tau+ 2 \mut \Phi'
\end{pmatrix}\,,\\
& =
\begin{pmatrix}
    -(m^t)^{-1}n &   (m^t)^{-1} \\
    (m^t)^{-1} & \ntil(m^t)^{-1}
\end{pmatrix}
\begin{pmatrix}
x_\sigma+ 2 \mus \Phi'\\
x_\tau+ 2 \mut \Phi'
\end{pmatrix}\,.
\end{align}
In a light cone basis these equations become
\begin{equation}
\xtil_\pm+2 \mu_\pm \Phitil''=\pm \,T_{\pm} \, \left(x_\pm+2 \mu_\pm
\Phi' \right) \,,\label{dual-1}
\end{equation}
where $T_{\pm}$ are orthogonal matrices given by
\begin{equation}
    T_{\pm} =(m^{t})^{-1}(I \mp n)\,.
    \label{eq:def-Tpm}
\end{equation}
Specifying to the case $n=0$ the above become
\begin{equation}
\xtil_\pm+2 \mu_\pm \Phitil''=\pm \,T \,(x_\pm+2 \mu_\pm \Phi') \,\label{h_du}
\end{equation}
with a single orthogonal matrix $T$.

A final curiosity is that $h$ according to eq.~\eqref{eq:h-2} is 
the Hodge dual of the respective term in the generating function 
\eqref{eq:h-1def}.

\section{Integrability Conditions}

Her we study the integrability conditions for the classical duality
equations.  It is instructive to study momentarily a more general
duality equation than \eqref{h_du}. Dimensional considerations
impose the form
\begin{equation}
\xtil_\pm^i+2\mu_\pm \util^i=\pm x_\pm^i \pm 2\mu_\pm u^i\,. \label{intcon_du}
\end{equation}
These equations are interpreted as equations on a bundle of
orthonormal coframes as in references
\cite{Alvarez:2000pk,Alvarez:2002mg}.  The vector valued functions
$u^{i}$ and $\util^{i}$ are functions on the same bundle.  We denote
by $'$ and $''$ we denote the derivatives with respect to $x$ and
$\xtil$.  Taking the derivative of \eqref{intcon_du} we have
\begin{align}
&\xtil^i_{+-} \mp x^i_{+-} -\omegatil^i{}_{j\mp}\xtil_\pm^j \pm \omega^i{}_{j\mp}x _\pm^j=\notag\\
&=2\mu_\pm\left[(-\util'^i{}_{;j}\pm u'^i{}_{;j})x^j_\mp+(-\util''^i{}_{;j}\pm u''^i{}_{;j})\xtil^j_\mp+
\omegatil^i{}_{j\mp}\util^j \mp \omega^i{}_{j\mp}u^j\right]+2\partial^2_{+-}\mu\left[-\util^i\pm u^i\right]\,.\label{intcon}
\end{align}

By the use of equations of motion \eqref{eom} we may eliminate second derivatives on the L.H.S of \eqref{intcon} which now reads as
\begin{equation}
-2\partial^2_{+-}\mu(\Phitil''^i\mp\Phi'^i)+\half ( \mp\Htil^i{}_{jk}\xtil_\pm ^j\xtil_\mp^k 
+H^i{}_{jk}x_\pm^j x_\mp^k ) -\omegatil^i{}_{j\mp}\xtil_\pm^j \pm \omega^i{}_{j\mp}x _\pm^j\,.
\end{equation}

Now the strategy is to use the duality equation \eqref{intcon_du} in
order to replace selectively\footnote{The subsequent equation may be
not therefore explicitly ``tilded-untilded'' symmetric, the final result
however has to be as neither $M$ nor $\Mtil$ is formally
distinguished.} \mbox{$\xtil_\mu^i$ with $x_\mu^i$.} The L.H.S is thus
\begin{align}
&-2\partial^2_{+-}\mu(\Phitil''^i\mp\Phi'^i)+\half ( \pm\Htil^i{}_{jk}+H^i{}_{jk})x_\pm^j x_\mp^k  \mp(\omegatil^i{}_{j\mp}
-\omega^i{}_{j\mp})x _\pm^j-4\mu_\pm\mu_\mp\Htil^i{}_{jk}\util^j u^k\\
&+\mu_\pm\Htil^i{}_{jk}(-\util^j \pm u^j)x^k_\pm+\mu_\mp
\Htil^i{}_{jk}(-\util^j \pm u^j)x^k_\pm-2\mu_\pm \omega^i{}_{j\mp}(-\util^j \pm u^j)\,.
%2(-\util'^i_j\pm u'i_j \pm \util''^i_j-u''i_j)\mu_\pm x^j_\mp+4\mu_pm\mu_mp(-\util''^i_j\pm u''i_j)(\util^j \pm u^j)+
%+2\partial^2_{+-}\mu(-\util^i \pm u^i)+2\mu_\pm(\omegatil^i_{j\mp}\util^j \mp \omega^i_{j\mp}u^j\)
\end{align}
Here, we have to identify as before
\cite{Alvarez:2000pk,Alvarez:2002mg} the orthogonal groups in both
coframes bundles by
\begin{equation}
\omegatil_{ij} + \half H_{ijk}\omegatil^{k} = \omega_{ij} + \half
\Htil_{ijk} \omega^{k}\,.\label{bunind}
\end{equation}
Using \eqref{bunind}, then again \eqref{intcon_du} and collecting the terms we obtain
\begin{align}
0=&2\partial^2_{+-}\mu(-\Phitil''^i\pm\Phi'^i+\util^i \mp u^i)+4\mu_\pm\mu_\mp(\util'^i{}_{;j}\mp u''^i{}_{;j})
(\util^j\pm u^j)\notag\\
&+\mu_\pm\left(\Htil^i{}_{jk}\util^k+H^i{}_{jk}u^k-2(\util'^i{}_{;j}\pm u'^i{}_{;j} \pm \util''^i{}_{;j} -u''^i{}_{;j})\right)x^j_\mp\notag\\
&+\mu_\mp\left(\mp H^i{}_{jk}(-\util^j\pm u^j)+\Htil^i{}_{jk}(-\util^j\pm u^j)\right)x^k_\pm\,.\label{intconf}
\end{align}
From $x^k_\pm$ and $x^j_\mp$ terms we learn respectively:
\begin{align}
H^i{}_{jk}u^k+\Htil^i{}_{jk}\util^k&=0 \, ,\label{heq1}\\
H^i{}_{jk}\util^k+\Htil^i{}_{jk}u^k&=0 \, ,\label{heq2}\\
\util'^i{}_{;j}+u''^i{}_{;j}&=0 \, ,\label{usec}\\
\util''^i{}_{;j}+u'^i{}_{;j}&=0 \, .\label{usec1}
\end{align}
From the first line of \eqref{intconf} and requiring $\mu$
independence leads to
\begin{equation}
u^i=\Phi'^i,\qquad\util^i=\Phitil''^i\label{intphi}
\end{equation}
and
\begin{align}
\util''^i{}_{;j}\util^j-u''^i{}_{;j} u^j&=0\,,\notag\\
u''^i{}_{;j}\util^j-\util''^i{}_{;j} u^j&=0\,.\label{intuc}
\end{align}
Here \eqref{intphi} tells that the integrable duality equation is \eqref{h_du}, the one
obtained already by studying the Hamiltonian formalism. In this case  
$u^i$ are only functions of $x$'s and $\util^i$ of $\xtil$'s, therefore mixed derivatives $u''^i{}_{;j}$ and
$\util'^i{}_{;j}$ vanish. Substituting \eqref{usec} and \eqref{usec1} into \eqref{intuc} and using \eqref{intphi}
we obtain
\begin{equation}
\Phi'_{i;j}\Phi'^j =\Phitil''_{i;j}\Phitil''^j 
\end{equation}
which corresponds to \eqref{phicond}, being expressed in a differential way. 

In the light-cone coordinates the trace of energy-momentum tensor is proportional to $T^{+-}$:
\begin{align}
\Tr\,[T^{\alpha\beta}] =e^{2\mu}\;T^{+-} &= 8e^{-2\mu}\left(\Phi'_{i;j}\,x^i_+x^j_-
+\Phi'_i \,x^i_{+-}\right), \\
&= 8e^{-2\mu}\left(\Phi'_{i;j}\,x^i_+x^j_-
-\half \Phi'_i H^i{}_{jk}\;x^j_+x^k_- 
-2 \Phi'_i \Phi'_{i}\;\partial^{2}_{+-}\mu(\vec{\sigma})\right).
\end{align}
In the last line the equation of motion was used.
Inspecting equations \eqref{heq1} and \eqref{heq2} we may establish
their connection to the relationship between $\Tr\,[T^{\alpha\beta}]$
and $\Tr\,[\Ttil^{\alpha\beta}]$.  Using the antisymmetry of
$H^i{}_{jk}$ together with \eqref{intphi} we write \eqref{heq1} and
\eqref{heq2} as
\begin{align}
\Phi'^i\,H^i{}_{jk}+\Phitil''^i\,\Htil^i{}_{jk}&=0 \, ,\label{heq3}\\
\Phitil''^i\,H^i{}_{jk}+\Phi'^i\,\Htil^i{}_{jk}&=0 \, .\label{heq4}
\end{align}
Now we contract the equation of motion \eqref{eom} with $\Phi'_i$ and obtain 
\begin{align}
\Phi'_i\,x^i_{+-}=-&\half \Phi'_i\, H^i{}_{jk}\;x^j_+x^k_--2\;\Phi'_i\,\Phi'^i\;\partial^{2}_{+-}\mu(\vec{\sigma})\notag\\
=&\half \Phitil''_i\, \Htil^i{}_{jk}\;x^j_+x^k_--2\;\Phitil''_i\,\Phitil''^i\;\partial^{2}_{+-}\mu(\vec{\sigma}).
\end{align}
In the last line we took advantage of \eqref{heq3} and \eqref{phicond}. 
Using the duality equation \eqref {intcon_du} we eliminate $x^j_\pm$'s in favor of their tilded counterparts. 
Having in mind the antisymmetry of $\Htil^i{}_{jk}$  
and \eqref{heq4} it is clear that the terms $\Htil^i{}_{jk}\,\Phi'^j\,\Phi'^k$, $\Htil^i{}_{jk}\,\Phitil''^j\,\Phi'^k$,
$\Htil^i{}_{jk}\,\Phitil''^j\,\Phitil''^k$ vanish. Hence we obtain
\begin{equation}
\Phi'_i\,x^i_{+-}=-\half \Phitil''_i\, \Htil^i{}_{jk}\;\xtil^j_+\xtil^k_-
-2\;\Phitil''_i\,\Phitil''^i\;\partial^{2}_{+-}\mu(\vec{\sigma})=\Phitil''_i\,\xtil^i_{+-}\,,
\end{equation}
which for the case $\Phi'_{i;j}=0$ is a statement that
$\Tr\,[T^{\alpha\beta}]$=$\Tr\,[\Ttil^{\alpha\beta}]$.  We understand
that equations \eqref{heq1} and \eqref{heq2} are a condition which
guarantees that \emph{form} of the trace of the energy-momentum tensor
is ``preserved on-shell'' by the duality transformation.

\section{Conclusions}
In order to set the duality equations
we have to impose the constraints allowing only the coupling of a curvature scalar to the dilaton fields whose 
differentials have the same norm in their respective metric. The special case is a linear dilaton field.
Here we might have expected to obtain a strong restriction, we required the ``conformal covariance'' as a local symmetry 
of Hamiltonian formalism and subsequently of the integrability conditions.

At the one-loop quantum level the condition\footnote{By imposing the
vanishing of beta function.} that the model has to satisfy in order to
be conformally invariant involves the second derivatives of a dilaton
field, \emph{e.g.}, \cite{GSW:Strings2,Polchinski:Strings1}.  It raises
therefore a question whether it is possible to establish at a quantum
level a more general form of duality transformation which leads to
preserving of the form of the beta function and what would be the role
of classical duality within such a construction.

\section*{Acknowledgments}

This work was supported in part by
National Science Foundation grants PHY--0244261 and PHY--0554821.

% \bibliographystyle{utphys}
% % Put in my bibliography
% \bibliography{oabib}

\providecommand{\href}[2]{#2}\begingroup\raggedright\endgroup

\end{document}